\documentclass[prd,preprint,superscriptaddress,nofootinbib,longbibliography]{revtex4}

\newcommand{\be}{\begin{equation}}
\newcommand{\ee}{\end{equation}}
\def\lta{\,\raise 0.3 ex\hbox{$ < $}\kern -0.75 em
 \lower 0.7 ex\hbox{$\sim$}\,}
\def\gta{\,\raise 0.3 ex\hbox{$ > $}\kern -0.75 em
 \lower 0.7 ex\hbox{$\sim$}\,}

\usepackage{xcolor}

\usepackage{graphicx}% Include figure files
\usepackage{dcolumn}% Align table columns on decimal point
\usepackage{bm,amsfonts,amsthm,amsmath,amssymb,array}
\usepackage{bbm}
\usepackage[]{epsfig,graphics}
\usepackage{comment}
\usepackage[T1]{fontenc}
\usepackage{lipsum}
\usepackage[utf8]{inputenc}
\usepackage{ulem}
\usepackage{multirow}
\usepackage{color}
\usepackage{lastpage}
\usepackage[sort&compress]{natbib}
\usepackage{enumerate}
\usepackage{wasysym}
\usepackage{relsize}
\usepackage{physics}
\usepackage{mathrsfs}
\usepackage{tensor}
\usepackage{xfrac}
\usepackage{siunitx}
\usepackage{setspace}  % helps with figure captions
  \usepackage[
         colorlinks=true,     
         urlcolor=blue,       % \href{...}{...} external (URL)
         ]{hyperref}
\usepackage{caption}
\usepackage{subcaption}
\def\ben{\begin{enumerate*}}
\def\een{\end{enumerate*}}
\def\bi{\begin{itemize*}}
\def\ei{\end{itemize*}}
\def\bd{\begin{description*}}
\def\ed{\end{description*}}
\def\be{\begin{equation}}
\def\ee{\end{equation}}
\def\bea{\begin{eqnarray}}
\def\eea{\end{eqnarray}}
\def\bfl{\begin{flushleft}}
\def\efl{\end{flushleft}}

\newcommand{\gsim}{\lower.7ex\hbox{$\;\stackrel{\textstyle>}{\sim}\;$}}
\newcommand{\lsim}{\lower.7ex\hbox{$\;\stackrel{\textstyle<}{\sim}\;$}}

\newcommand{\beq}{\begin{equation}}
\newcommand{\eeq}{\end{equation}}

\begin{document} 

\title{Fluctuations in Hill's equation parameters and application to cosmic reheating}

%\title{Generalized Models for Inflationary Preheating:\\ %Oscillations with Quartic Effective Potentials}

\author{Leia Barrowes}
\email{barrowes@umich.edu}
\affiliation{Physics Department, University of Michigan, Ann Arbor, MI 48109, USA} 

\author{Fred C. Adams}
\email{fca@umich.edu}
\affiliation{Physics Department, University of Michigan, Ann Arbor, MI 48109, USA} 
\affiliation{Astronomy Department, University of Michigan, Ann Arbor, MI 48109, USA}

\author{Anthony M. Bloch}
\email{abloch@umich.edu}
\affiliation{Mathematics Department, University of Michigan, Ann Arbor, MI 48109, USA}

\author{Scott Watson}
\email{gswatson@syr.edu}
\affiliation{Department of Physics, Syracuse
  University, Syracuse, NY 13244, USA} 

\begin{abstract}
Cosmic inflation provides a compelling framework for explaining several observed features of our Universe, but its viability depends on an efficient reheating phase that converts the inflaton's energy into Standard Model particles. This conversion often proceeds through non-perturbative mechanisms such as parametric resonance, which is described by Hill’s equation. In this work, we investigate how stochastic fluctuations in the parameters of Hill's equation can influence particle production during reheating. We show that such fluctuations can arise from couplings to light scalar fields, and can significantly alter the stability bands in the resonance structure, thereby enhancing the growth of fluctuations and broadening the region of efficient energy transfer. Using random matrix theory and stochastic differential equations, we decompose the particle growth rate into deterministic and noise-induced components and demonstrate analytically and numerically that even modest noise leads to substantial particle production in otherwise stable regimes. These results suggest that stochastic effects can robustly enhance the efficacy of reheating across a wide swath of parameter space, with implications for early-Universe cosmology, UV completions involving multiple scalar fields, and the resolution of the cosmological moduli problem.

\end{abstract}

\maketitle

\section{Introduction}
\label{sec:intro}

The paradigm of cosmic inflation, a period of accelerated expansion that set the initial conditions for our observable Universe, is strongly supported by observations of the Cosmic Microwave Background (CMB), which reveal a nearly scale-invariant power spectrum of primordial perturbations~\cite{Guth1981, Bardeen83, plank2018inflation}. To smoothly exit this inflationary phase, in the standard paradigm, the inflaton field must undergo coherent oscillations about the minimum of its potential, facilitating the transition to an eventual radiation-dominated Universe suitable for Big Bang Nucleosynthesis (BBN)~\cite{AS_newinfl, Linde_newinfl}. These oscillations naturally open the possibility for non-perturbative energy transfer mechanisms, known as (p)reheating~\cite{Traschen:1990sw,Kofman94, Kofman:1997yn}.

In this work, we study how random fluctuations affect particle production during non-perturbative reheating. While quantum fluctuations are an intrinsic aspect of any field-theoretic treatment, we focus on more substantial fluctuations arising from a collection of light scalar fields---such as compactification moduli, axions, or excitations from brane dynamics---which can introduce significant stochastic effects into the reheating process.\footnote{Although we invoked scalar fields as the source of fluctuations in this work, our approach can be used to include fluctuations from any source.} 

The role of stochasticity in reheating, both homogeneous~\cite{Zanchin1998} and inhomogeneous~\cite{Zanchin_1999}, has been explored in previous studies. These previous efforts used results from the theory of random matrices~\cite{furstenberg_random_matrices} to show that noise generally enhances particle production. More broadly, the growth rate of the instability can be decomposed into an average term plus a positive-definite random walk contribution~\cite{AdBloch2008}, a decomposition that we will revisit and extend in this paper for the case of spatially homogeneous noise.

To systematically investigate the influence of fluctuations, we adopt the random transfer matrix approach and analyze the resulting stochastic Hill's equations in the context of preheating. We find that even modest fluctuation levels destabilize regions of parameter space that would otherwise be broadly or narrowly stable. Although broad resonance bands are slightly suppressed, the overall effect is an expansion of the region where non-perturbative reheating occurs. This result underscores the ubiquity of preheating in a landscape with stochastic perturbations.

Several complementary analytical techniques support our findings. In the narrow resonance regime with weak noise, the dynamical renormalization group reveals that stochasticity can be recast as an effective interaction term that modifies correlation functions~\cite{murakami2016nonperturbativeevaluationquantumparticle}. We recover similar terms and generalize the result across parameter space. An analogy with Anderson localization in disordered media~\cite{anderson_loc} allows one to interpret the Lyapunov exponent as a proxy for particle number growth, and the associated methods from condensed matter physics yield a Fokker-Planck equation governing the evolution of the particle distribution~\cite{Amin2015FromWT, Amin2017MultifieldSP}.

The ultraviolet (UV) origin of these fluctuation-inducing light spectator fields, and their potential disruptive effects on observable quantities such as curvature perturbations, have been studied in~\cite{Gu_2020}, with proposed strategies for mitigating such effects. Indeed, UV theories suggest the presence of multiple scalar fields, and these can impact both inflation and reheating, and can potentially offer solutions to the Hierarchy Problem (see, e.g. \cite{Easther:2005zr,Arkani-Hamed:2016rle}). In this paper, we generalize the study of these implications and offer new insights into the dynamics of reheating in an environment where fluctuations are significant. 

These findings also have implications for Effective Field Theory (EFT) approaches to inflation~\cite{weinberg_EFTinfl, Cheung:2007st} and reheating~\cite{Ozsoy:2014sba, Ozsoy:2017mqc}. In particular, the inclusion of stochastic source terms, as proposed in~\cite{LopezNacir:2011kk, Ozsoy:2017mqc}, can help capture nonlinearities and broaden the parameter space of viable reheating scenarios. Moreover, noise-induced instabilities may provide a new mechanism to address the cosmological moduli problem~\cite{Kane:2015jia}, especially where parametric resonance is otherwise constrained by fine-tuned couplings~\cite{2017-gauge-CMP, deal2025cosmologicalmodulinonperturbativeproduction, Shuhmaher_2006}.

This present work employs random matrix and stochastic methods to analyze the effects of fluctuations on reheating, and demonstrates that randomness enhances the efficacy of non-perturbative particle production. Our approach complements EFT treatments and provides a bridge between high-energy UV model-building and non-equilibrium statistical dynamics in the early Universe. The rest of the paper is organized as follows. Section \ref{sec:parametric} reviews the standard method for finding characteristic growth exponents associated with parametric resonance. Section \ref{sec:fluctuatehill} introduces random fluctuations into the problem and shows how they affect these growth rates. Additional details are provided in Appendix \ref{sec:stable-flucs}, which derives an analytic form for the growth rate for regions of parameter space that are stable in the absence of fluctuations. Section \ref{sec:fluctuating-fields} illustrates how we might achieve fluctuations in a reheating scenario from a collection of light scalars, and the (non-lattice) numerical results in Section \ref{sec:results} demonstrate their success in enhancing reheating.

\section{Parametric Resonance}
\label{sec:parametric}

In this section, we start with a brief review of Hill's equation without fluctuations to define notation and illustrate the stability diagram that determines the conditions required for resonance to occur. Consider the general form of Hill's equation written in the form 
\be
{\ddot u} + \Omega^2 u = 0 \,,
\ee
where $\Omega^2$ is a periodic function with period $T$, so that $\Omega^2(t+T) = \Omega^2(t)$. For example, the Mathieu equation in one standard form has $\Omega^2(t)=A - 2q \cos2t$. According to Floquet's theorem \cite{ASENS_1883_2_12__47_0}, solutions take the form
\begin{align}
    u(t) = e^{\mu t} P_1(t) + e^{-\mu t} P_2(t)
\end{align}
where $P_1$ and $P_2$ are periodic functions and $\mu$ is the Floquet exponent, whose real part is the Lyapunov exponent $\gamma=\text{Re}(\mu)$, which we colloquially refer to as the growth rate.
On the interval $[0,T]$, the equation has two linearly independent solutions. We choose these two solutions to be the principal solutions, which are defined such that  
\be
u_1(0) = 1 \qquad {\rm and} \qquad {\dot u}_1 (0) = 0
\ee
and 
\be
u_2(0) = 0 \qquad {\rm and} \qquad {\dot u}_2 (0) = 1\,.
\ee
Any solution can be written as a linear combination  of the two principal solutions. We conceptually break up the time evolution into intervals of length $T$. For a given interval, labeled here by the index $n$, the solution has the form 
\be
u_n(t) = \alpha_n u_1(t) + \beta_n u_2(t) \,. 
\ee
Similarly, for the next interval, labeled as $n+1$, we have
\be
u_{n+1}(t) = \alpha_{n+1} u_1(t) + \beta_{n+1} u_2(t) \,. 
\ee
If we enforce continuity of both the function and its derivative at the interface between successive cycles, we obtain two matching conditions:
\be
\alpha_n u_1(T) + \beta_n u_2(T) =
\alpha_{n+1} u_1(0) + \beta_{n+1} u_2(0) = \alpha_{n+1} 
\ee
and 
\be
\alpha_n {\dot u}_1(T) + \beta_n {\dot u}_2(T) =
\alpha_{n+1} {\dot u}_1(0) + \beta_{n+1} {\dot u}_2(0) = \beta_{n+1} 
\ee
These two results thus take the form
\be
\begin{bmatrix}
    \alpha_{n+1}  \\ \beta_{n+1} 
\end{bmatrix} =
\begin{bmatrix}
u_1(T)  & u_2(T) \\
{\dot u}_1(T) & {\dot u}_2(T) 
\end{bmatrix}
\begin{bmatrix}
    \alpha_n \\
    \beta_n
\end{bmatrix} 
\equiv \mathbb{M}_n 
\begin{bmatrix}
    \alpha_n \\
    \beta_n
\end{bmatrix} \,,
\label{eqn:uM}
\ee
where the final equality defines the transfer matrix $\mathbb{M}_n$. Standard arguments \cite{MagWink1966} show that the determinant of the transfer matrix must be unity, so that the number of independent matrix elements is reduced from four to three. In addition, {\it for the case where the equation is symmetric with respect to the midpoint of the time interval}, we have the additional condition $u_1(T)={\dot u}_2(T)$.\footnote{Note that much of the literature on preheating does not take into account the symmetry condition $u_1(T)={\dot u}_2(T)$. This condition provides a useful consistency constraint on the numerical evaluation of the matrix elements, and can be used to simplify expression for the eigenvalues. }  As a result, only two of the matrix elements are independent. Here we define 
\be
h_n \equiv u_1(T) \qquad {\rm and} \qquad
g_n \equiv {\dot u}_1(T) \,,
\label{hgdefine} 
\ee
so that the transfer matrix has the form
\be
\mathbb{M}_n =
\begin{bmatrix}
     h_n & (h_n^2-1)/g_n \\
     g_n & h_n 
\end{bmatrix} \,. 
\label{eqn:transfer} 
\ee

If all of the cycles are the same, which will be the case in the absence of stochasticity, then the matrix elements will be the same for all cycles, and the eigenvalues are given by 
\be
\lambda = h \pm \sqrt{h^2 -1}\,.
\ee 
After $N$ cycles, the solution to the differential equation is then given by the form 
\be
\begin{bmatrix}
    \alpha_N \\
    \beta_N 
\end{bmatrix}
= \mathbb{M}^N 
\begin{bmatrix}
    \alpha_0 \\ 
    \beta_0 \,.
\end{bmatrix} 
\label{ncycle} 
\ee
We can rewrite the vector $[\alpha_0,\beta_0]$ in the form
\be
\begin{bmatrix}
    \alpha_0 \\
    \beta_0 
\end{bmatrix}
= A {\vec V}_1 + B {\vec V}_2 \,,
\ee
where the ${\vec V}_i$ are the eigenvectors of the matrix $\mathbb{M}$, corresponding to eigenvalues $\lambda_i$. Without loss of generality, we can take the largest eigenvalue to be $\lambda_1$. After $N$ cycles, the matrix/vector form of the solution becomes 
\be
\begin{bmatrix}
    \alpha_N \\ 
    \beta_N 
\end{bmatrix} \approx
\lambda_1^N A {\vec V}_1 \,.
\ee
If we take the limit $N\to\infty$, the growth rate for the solution is found to be 
\be
\gamma = {1 \over T} \log\lambda_1 \,. 
\label{eqn:gama-loglam}
\ee

\section{Effects of Fluctuations on the Instability Diagram}
\label{sec:fluctuatehill}

This section introduces fluctuations into our consideration of Hill's equation. Note that fluctuations can be included in two related but different ways. [A] The parameters that appear in Hill's equation can vary from cycle to cycle.
For example, in the case of Mathieu's equation, the frequency and forcing strength $(A,q)$ become $(A_n,q_n)$, where $n$ labels the cycle and the parameters are sampled from a distribution. [B] The Hill's equation can be generalized to include at least one additional term that is a stochastic process. These two approaches are related, in that the integral of the stochastic process in case [B] produces the distribution of parameters for case [A] (for further detail, see Ref. \cite{Adams_2013} and Appendix A). In this work we consider the cycle to cycle variations of case [A].

We thus consider the case where the parameters of the Hill's equation, and hence the solutions, vary from cycle to cycle. In this case, the evolution after $N$ cycles is given by the generalization of equation (\ref{ncycle}), i.e., 
\be
\begin{bmatrix}
    \alpha_N \\ 
    \beta_N 
\end{bmatrix} = 
\left\{ \prod_{n=1}^N \mathbb{M}_n \right\} 
\begin{bmatrix}
    \alpha_0 \\
    \beta_0 
\end{bmatrix}\,.
\label{gencycle} 
\ee
In the presence of fluctuations, the matrix elements in the $\mathbb{M}_n$ will be different in each cycle, where the entries depend on how the fluctuations determine the distribution of parameters appearing in Hill's equation. In any case, with nonzero fluctuations, the growth rate generalizes to the form
\be
\gamma T = \lim_{N\to\infty} \log \left\Vert 
\prod_{n=1}^N \mathbb{M}_n \right\Vert \,,
\label{eqn:gama-normM}
\ee
where the vertical bars denote taking the norm, but the result is independent of what norm is used \cite{Furstenberg1960,Furstenberg1963}.

\begin{figure}[t]
\includegraphics[scale=0.50]{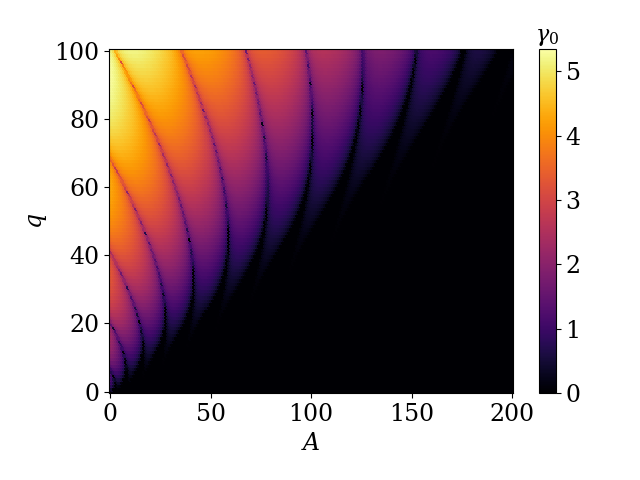}
\includegraphics[scale=0.5]{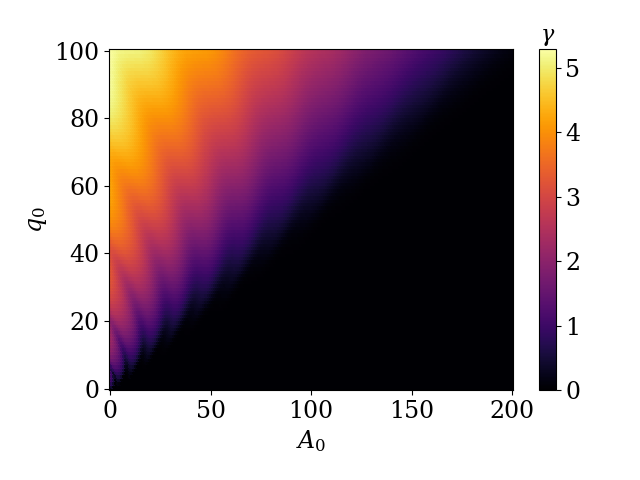}
% \vskip-3.0truein
\caption{Left: Floquet chart in $(A,q)$ parameter space for the Mathieu equation without fluctuations.
Right: Floquet chart in $(A_0, q_0)$ parameter space with fluctuations included; for all $j$ in $N=10^5$ cycles, $\xi_j$ was selected from a normal distribution with $\sqrt{\text{Var}\ \xi}=10^{-1}$, with the parameters in the Mathieu equations given by $A_j=A_0/\left(1+\xi_j\right)$ and $q_j=q_0/\left(1+\xi_j\right)$. } 
\label{fig:quartplane} 
\end{figure}

\begin{figure}[t]
\includegraphics[scale=0.50]{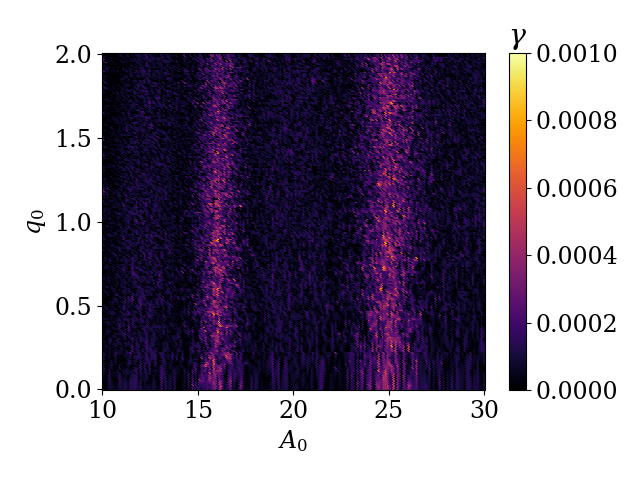}
\includegraphics[scale=0.5]{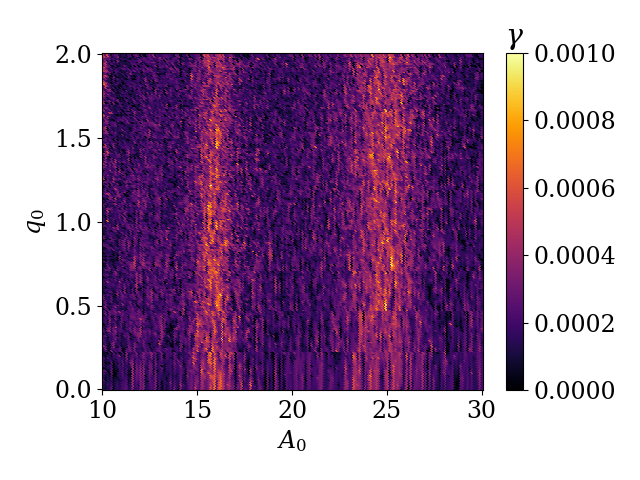}
% \vskip-3.0truein
\caption{Floquet chart in $(A_0, q_0)$ parameter space with fluctuations. For all $j$ in $N=10^5$ cycles, $\xi_j$ was selected from a normal distribution with $\sqrt{\text{Var}\ \xi}=5\times10^{-2}$ (right) and $10^{-1}$ (left), with the Mathieu parameters given by  $A_j=A_0/\left(1+\xi_j\right)$ and $q_j=q_0/\left(1+\xi_j\right)$. Without fluctuations, growth rates in this region are essentially zero (with $\gamma_0\lesssim10^{-16}$ in the numerical treatment).
There are still some bands of stability in this previously completely stable region where the approximations in Appendix \ref{sec:stable-flucs} break down \cite{AdBloch2009}, but the bands of instability are not insignificant and become broader with increasing amplitude.} 
\label{fig:lowq-charts} 
\end{figure}

For example, one way to take the norm is to start with an arbitrary vector $(\alpha_0,\beta_0)$ as in equation (\ref{gencycle}), and then use the usual Euclidean norm at the $N$th step. In this case, the expression for the growth rate becomes 
\be
\gamma T = \lim_{N\to\infty} {1\over N} 
\log \left[(\alpha_N^2 + \beta_N^2)^{1/2} \right] \,.
\label{limit}
\ee

Taking the limit in equation (\ref{limit}) is equivalent to finding the eigenvalue of an infinite product of matrices, where the matrix elements are chosen from a well-defined distribution. The matrix elements themselves are the solutions to Hill's equation for a particular cycle, where the parameters $(A_n,q_n)$ are chosen from well-defined distributions. Because the eigenvalue of the product of the matrices is {\it not} the product of the individual eigenvalues \cite{Furstenberg1960,Furstenberg1963,LimaRahibe}, the evaluation of the growth rate is complicated (see also \cite{AdBloch2008,AdBloch2009,AdBloch2010}).

The growth rate for this scenario of matrix products has been considered previously. In the unstable limit, where the eigenvalues of the individual matrices are primarily in the growing (unstable) regime, the matrix elements are mostly in the regime $|h_n|>1$. In this unstable limit, a particular sample of the parameters could result in a point $(A_n,q_n)$ that lies a band of stability (where $|h_n|<1$). However, the bands of stability have exponentially shrinking width in the limit of large $q$, so that any finite fluctuation amplitude will be larger than the width for sufficiently large $q$. In any case, one can write the transfer matrix from Eqn. (\ref{eqn:transfer}) in the form 
\be
\mathbb{M}_n = h_n 
\begin{bmatrix}
     1 & h_n/g_n - 1/(g_n h_n) \\
     g_n/h_n & 1 
\end{bmatrix} 
= h_n \mathbb{C}_n \,, 
\label{transfertwo} 
\ee
where the second equality defines the matrix $\mathbb{C}_n$. This separation can be translated into separating the growth rate into two pieces \cite{AdBloch2008}, 
\be
\gamma = \gamma_\infty + \gamma_R \,. 
\label{eqn:gamma-parts}
\ee
The first part, the asymptotic growth rate, is given by 
\be\
\gamma_\infty = \lim_{N\to\infty}
{1\over NT} \sum_{n=1}^N \log (2h_n) \,, 
\ee
which corresponds to an average of the growth rates for the individual cycles. The second part $\gamma_R$ of the growth rate arises due to the stochastic growth from multiplication of the reduced matrices $\mathbb{C}_n$. In the highly unstable limit, where $|h_n|\gg1$, the random component $\gamma_R$ can be evaluated in closed form \cite{AdBloch2008}, although this regime often has $\gamma_R\ll\gamma_\infty$. In the stable limit, where $|h_n|<1$ and $\gamma_\infty=0$, the growth rate is dominated by $\gamma_R$, and its calculation, originally from \cite{AdBloch2009}, is shown in Appendix \ref{sec:stable-flucs} along with a numeric demonstration of its validity in Fig. \ref{fig:ellpt-numeric}. The finding that $\gamma_R>0$ in (formerly) stable regions is the major result that we wish to highlight and utilize in the preheating context---this nonzero growth rate can nudge reheating along in previously unviable regions of parameter space and thereby increase the range of feasible inflation models.

Figure \ref{fig:quartplane} shows Floquet charts with and without fluctuations for a range of the $A-q$ plane. The upper left region for $2q>A$ shows the characteristic bands of stability and instability for the standard Mathieu equation (left panel). The stability bands disappear with the inclusion of fluctuations (right panel). In this region we have $\gamma\approx\gamma_\infty$, where the dominant effect of fluctuations is the local averaging of non-fluctuating growth rates, as illustrated by the blurring of features in the color map. Figure \ref{fig:lowq-charts} zooms in on a broad stability region, $A\gtrsim2q$, where the non-fluctuating growth rates are effectively zero, meaning $\gamma_\infty$ also vanishes. As a result, the non-zero values shown in these figures are dominated by growth due to fluctuations with $\gamma\approx\gamma_R>0$. A residual band structure still appears in some regions of parameter space, with some growth rates higher than others. In any case, the nonzero growth rate found here is the main mathematical result of interest. This finding implies that fluctuations can drive exponential growth in otherwise stable regions of parameter space, and this generalization can be a useful tool for ensuring successful reheating of the universe after inflation.

\section{Application to Cosmic Reheating}

Cosmic inflation in its simplest realization is driven by a scalar field $\phi$ with a vacuum energy that accelerates the expansion of the Universe. The reheat field $\chi$ is diluted to its vacuum state long before inflation ends, and we wish to convert energy into the $\chi$ fields through the reheating process.
The action of these two scalars can be written in the form
\begin{align}
    S = \int \text{d}^4x\sqrt{-g} \left(\frac{m_\text{pl}^2}{2}R +\frac{1}{2}\partial_\mu\phi\partial^\mu\phi +\frac{1}{2}\partial_\mu\chi\partial^\mu\chi - V(\phi,\chi) \right)
    \label{eqn:action}
\end{align}
where
\begin{align}
    V(\phi,\chi)=\frac{m_\phi^2}{2}\phi^2 + \frac{m_\chi^2}{2}\chi^2 + \frac{\lambda}{4}\phi^4 + \frac{\lambda_\chi}{4}\chi^4 + V_\text{int}
    \label{eqn:Vint}
\end{align}
These inflaton potential terms don't necessarily reflect the shape of the potential during inflation;
rather, we expand the potential around the coherently oscillating minimum where reheating occurs.
We include the quartic term for several reasons: in the case of high inflaton potential amplitude $\Phi$, this quartic term may become comparable to the mass term.
Secondly, it is possible for there to be no mass at all, such as the minima of the following common inflaton potentials, 
\be
V(\phi) = {1\over2} \Lambda^4 \left( 1 + \cos\frac{\phi}{f} \right)^2 \qquad {\rm or} \qquad V(\phi) = \Lambda^4 
\left[1 - \left(\frac{\phi}{f}\right)^2 \right]^4 \ .
\ee
Most generally, both terms will be present, at least from quantum corrections.

At this point we have not yet introduced fluctuations. We will demonstrate the non-fluctuating behavior from this potential before showing the additional effects of stochasticity in the next section. Here we illustrate the mechanism using the potential $V_\text{int}=\sigma \phi\chi^2$, although other choices are possible with similar proceeding calculations \cite{Dufaux_2006, paper1}.
This particular interaction term requires $\lambda_\chi>2\sigma^2/m_\phi^2$ to bound the potential from below. Besides ensuring physical solutions, however, this quartic term plays a negligible role in the dynamics of both fields. The remaining parameters, $m_\phi, m_\chi, \lambda \geq0$ may or may not be present for the following analysis, where we will apply the methods of Sections \ref{sec:parametric} and \ref{sec:fluctuatehill} to the equations of motion of Eqns. (\ref{eqn:action}) and (\ref{eqn:Vint}).

The equation of motion for the homogeneous inflaton field is given by 
\begin{align}
    \ddot \phi + m_\phi^2\phi + \lambda \phi^3=0\ ,
    \label{eqn:phi-EOM}
\end{align}
where we neglect backreaction from both $\chi$ and self-interactions. This approach is valid considering the dominance of the inflaton's energy \cite{paper1}, although such interactions are important for thermalizing the Universe at the end of reheating \cite{Kofman94}.

\begin{figure}
    \centering
    \includegraphics[width=0.49\linewidth]{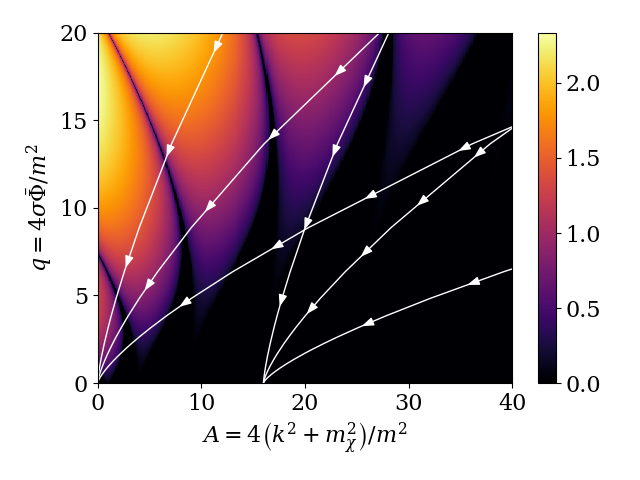}
    \includegraphics[width=0.49\linewidth]{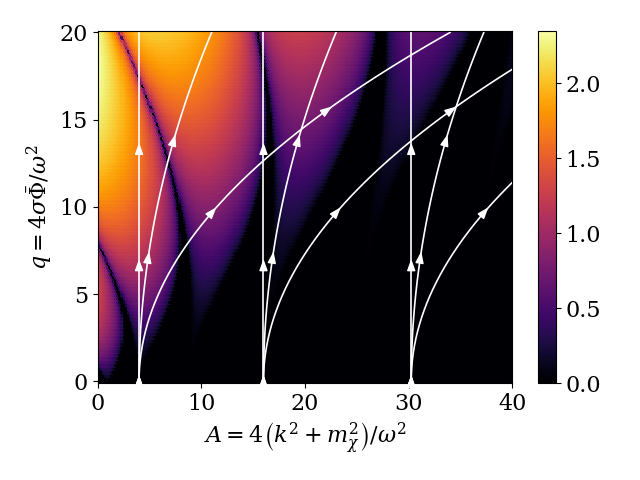}
    \caption{Stability maps and expansion trajectory examples for $V(\phi)=\frac{1}{2}m_\phi^2\phi^2$ (left) and $V(\phi)=\frac{1}{4}\lambda\phi^4$ (right). The massive potential trajectories always have decreasing $q$ and the quartic potential trajectories always have increasing $q$. Notice the slight difference in the underlying colormap shape from $\epsilon\neq0$ on the right.}
    \label{fig:traj-maps}
\end{figure}

The general solution to Eqn. (\ref{eqn:phi-EOM}) is written in terms of elliptical integrals, but is very well approximated by \cite{paper1}
\begin{align}
    \phi(t)=\Phi(t)\cos(\omega t)\left[1-\epsilon\sin^2(\omega t)\right]
\end{align}
where the oscillation frequency $\omega$ varies with expansion according to,
\begin{align}
    \omega=\frac{\left(m_\phi^2+\lambda\Phi^2\right)^{3/2}}{m_\phi^2+\frac{9}{8}\lambda\Phi^2} + \dots
    \ \to\ \begin{cases}
        m_\phi &\text{ for $\lambda\to 0$} \\
        \tilde\omega\sqrt{\lambda}\Phi &\text{ for $m_\phi^2\to0$}
    \end{cases}
    \label{eqn:omega}
\end{align}
where $\tilde\omega=2\pi^{3/2}\Gamma^{-2}\left(\tfrac{1}{4}\right)\approx 0.85$.
The parameter $\epsilon$ captures the anharmonic aspect of the inflaton's oscillation from the quartic term,
\begin{align}
    \epsilon = \left(9+8m_\phi^2/\lambda\Phi^2\right)^{-1}+\dots
    \ \to\ \begin{cases}
        0 & \text{ for $\lambda\to0$} \\ 8\sqrt{2}\ \tilde\omega \sech\frac{3\pi}{2} &\text{ for $m_\phi^2\to0$}
    \end{cases}\ .
\end{align}
We appropriately approximate $\Phi$ to be constant for each oscillation.
However, $\Phi$ varies adiabatically, not stochastically, from cycle to cycle because of the expansion of the Universe with scale factor $a$ \cite{turner-COs},
\begin{align}
    \Phi\propto a^{3(1+w)/p}\ , \quad\text{where}\quad w=\frac{p-2}{p+2} \quad\text{and}\quad V(\phi)\propto \phi^p
    \label{eqn:general-w}
\end{align}
for a monomial power law $V(\phi)$ with power $p$, which is generalized to a slowly-varying equation-of-state for the binomial $V(\phi)$ in Eqn. (\ref{eqn:Vint}),
\begin{align}
    w = \frac{\lambda\langle\phi^2\rangle_T}{4m_\phi^2\langle\phi^2\rangle_T + 3\lambda\langle\phi^4\rangle_T}
    \quad \to
  \quad \Phi \propto \begin{cases}
        a^{-3/2} &\text{ for $\lambda\to0$} \\a^{-1 } &\text{ for $m_\phi^2\to0$}
    \end{cases}
\end{align}
The solution of $\phi$ enters into the EOM of $\chi$'s momentum modes, which can be written as a Hill's equation with parameters $A$ and $q$,
\begin{gather}
    \ddot\chi_k + \left[k^2+m_\chi^2+2\sigma\phi(t)\right]\chi_k=0\\
\begin{split}
    \chi_k'' + \left(A+2q\cos(2\tau)\left[1-\epsilon\sin^2(2\tau)\right]\right)\chi_k=0 \\
    \text{where}\quad
    A=\frac{4(k^2+m_\chi^2)}{\omega^2}\ ,\quad 
    q = \frac{4\sigma\Phi}{\omega^2}\ ,\quad
    \tau=\frac{\omega t}{2}
\end{split} \label{eqn:hills}
\end{gather}
% We see that we recover all that is familiar from Sec. \ref{sec:fluctuating-fields} where $p=2$ only.
The most consequential difference for $p=4$ is codified in Eqn. (\ref{eqn:omega}), which indicates the decrease in oscillation frequency with inflaton amplitude. This behavior results in the $q$-parameter \textit{increasing} with expansion rather than decreasing like the case with $p=2$.  Figure \ref{fig:traj-maps} shows some example trajectories with different wavenumbers $k$ and different values of $m_\chi^2$. For $p=2$, all trajectories tend toward $\gamma=0$ except for an interesting case where $m_\phi$ is slightly greater than $m_\chi/2$, which is simply the kinematic preference in the perturbative limit $q\ll1$. In contrast, the $p=4$ trajectories all lead to increasing $q$. When $m_\chi=0$, the $A$-parameter is constant, resulting in an indefinitely increasing growth rate that guarantees complete reheating. However, any $m_\chi^2>0$ makes the trajectories veer quadratically towards decreasing $A$; as a result, even with an initial period of increasing growth rate, the trajectory will inevitably approach the region of broad stability, $A\gsim2q$. For a general mixed potential, the trajectories will start off with increasing $q$ before turning back toward decreasing $q$ once more; the quartic behavior dominates for high $\Phi$ and the quadratic potential behavior dominates for low $\Phi$ \cite{paper1}.
We see that in the majority of these cases, we are in danger of incomplete reheating by ending up in a broad stability part of parameter space.
Stochasticity in the equations of motion increase the efficacy of preheating by making $\gamma>0$ in these otherwise barren regions.
Higher wavenumbers $k$ of $\chi$, higher reheat field mass $m_\chi$, smaller coupling $\sigma$, and lower inflaton amplitude $\Phi$ all contribute to the possibility of broad stability; in particular, if $2\Phi\sigma\lesssim k^2+m_\chi^2$ then the classical growth rate vanishes.

\begin{figure}
    \centering
    \includegraphics[width=0.7\textwidth]{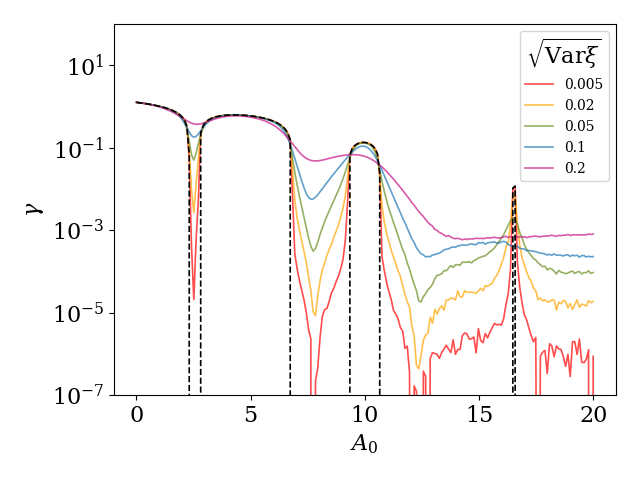}
    \caption{Growth rates vs $A_0$ for $q_0=4$ with no fluctuations (dashed black) and with several choices of $\sqrt{\text{Var}\ \xi}$ (solid). We see increases in growth rate in the stability bands and slight decreases in growth rate in the broad resonance bands as expected from Eqn. (\ref{eqn:gamma-parts}). Some numeric uncertainty is visible in broad stability regions.}
    \label{fig:high-q-lines}
\end{figure}

Crucially, random fluctuations would lift these growth rates to be small but no longer zero, as shown in Appendix \ref{sec:stable-flucs} and Fig. \ref{fig:quartplane}, \ref{fig:lowq-charts} and \ref{fig:high-q-lines}.
High-wavenumber modes in particular can experience growth that could otherwise have been impossible, blueshifting the resulting reheat field spectrum and giving us more total reheat field energy.

Small values of the coupling result in complete reheating over the course of one to a few oscillations of the inflaton \cite{paper1}, so that $(A,q)$ are effectively stationary and $\Phi$ has a single value before going to zero as the inflaton's energy is transferred almost immediately into the reheat field.
Such values for the coupling are motivated by the approaches of string theory to reheating and were studied in a numerical approach in \cite{Braden:2010wd}. We also note that even smaller values of this coupling could result from spontaneously broken symmetries. Our work here differs from \cite{Braden:2010wd} as we both analytically and numerically calculate trajectories through the parameter space.
While a large value of $\sigma$ generally results in successful reheating, a small $\sigma$ is also possible, in which case the inflaton oscillates many times before transferring its energy away, if ever.
In this case we must consider the expansion trajectories and the fact that exponential growth rate is more affected by fluctuations, which would give hope of successful reheating despite the smallness of $\sigma$.

\section{Fluctuations during Reheating} \label{sec:fluctuating-fields}

\begin{figure}[h]
    \centering
    \includegraphics[width=0.49\linewidth]{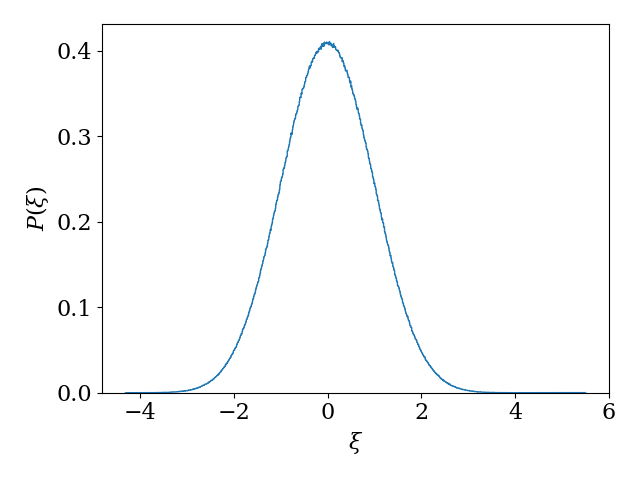}
    \includegraphics[width=0.49\linewidth]{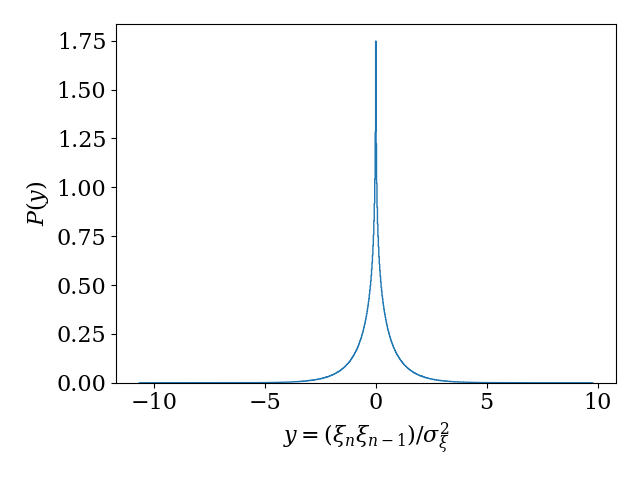}
    \caption{
    This example shows a distribution of $\xi$ (right) at the end of each $\phi$-oscillation for $L=20$ ancillary fields $\psi_i$ with masses uniformly selected from $m_i\in(0, m_\phi)$ that evolve like Eqn. (\ref{eqn:psi-cos}). The combination $g_i \Psi_i\Lambda_\psi/m_\phi^2$ is uniformly selected from $g_i \Psi_i\Lambda_\psi/m_\phi^2 \in (0, 2/\sqrt{L})$; the factor of $1/\sqrt{L}$ normalizes the variance to ${\text{Var}\ \xi}\approx1$ for any $L$.
    Crucially, $P(\xi)$ is a normal distribution, and we find this to be the case for $L\geq6$. 
    The correlation between periods $n$ (left) is centered close enough to zero, $|\langle y\rangle|\sim 10^{-2}$, that we can consider consecutive samples of $\xi$ to be uncorrelated. }
    \label{fig:xi-dist-numeric}
\end{figure}

To make the above notions more concrete, we next want to induce fluctuations in the inflaton that will result in cycle-to-cycle variations in the reheat field's EOM.
If the inflaton $\phi$ is coupled to many other scalar fields $\psi_i$ by
\begin{align}
    V_\text{int} (\phi,\psi_i) = \frac{\Lambda_\psi}{2}\sum_i^L g_i \phi^2 \psi_i
    \label{eqn:psi-int}
\end{align}
where $\Lambda_\psi$ is some characteristic energy scale, then the EOM of the inflaton becomes, sticking with the illustrative $p=2$ case,
\begin{align}
    \ddot\phi + m_\phi^2\left(1 + \xi \right)\phi=0 \qquad \text{where} \qquad \xi\equiv \frac{\Lambda_\psi}{m_\phi}\sum_ig_i\frac{\psi_i}{m_\phi}\ .
\end{align}
Scalar fields are abundantly predicted when considering reductions of higher-energy and higher-dimensional theories.
In our demonstrative analysis we use $\xi\lesssim 1$, which loosely constrain $\Lambda_\psi$ and the magnitude of $\psi_i$.
One might wonder about whether $\phi$ will cause growth in $\psi_i$ similarly to its resonant effect on $\chi$.
The particular choice of interaction in Eqn. (\ref{eqn:psi-int}) results in linear growth in the $\psi_i$ fields, which is subdominant to the prospective exponential growth in the $\chi$ field.
Other options such as $V_\text{int}=\frac{1}{2}\sum_i g_i\phi^2\psi_i^2,\ V_\text{int}=\frac{1}{2}\Lambda_\psi\sum_i g_i\phi\psi_i^2$ would result in exponential growth in $\psi_i$.
These effects are not computationally included in our demonstration of the efficacy of fluctuations in augmenting preheating, but qualitatively any $\psi_i$ growth would serve to intensify the desired effect: extra fields coupled to $\phi$ transfer energy more effectively and would make reheating complete sooner; in addition, the increased $\psi_i$ amplitudes would increase the size of the fluctuations in the $\chi$-EOM, enhancing the growth rate increase resulting from stochasticity. As a result, by neglecting of the interaction in Eqn. (\ref{eqn:psi-int}) on the $\psi_i$ EOMs, our result provides a conservative estimate of the advantage that stochasticity provides to particle production efficiency.
We will defer a full numerical analysis of the system of $\phi$, $\chi$, $\psi_i$ to future work.
In our non-lattice numerical results following this section, we will simply use
\begin{align}
    \psi_i = \Psi_i \cos(m_i t) \ ,
    \label{eqn:psi-cos}
\end{align}
which results in $\xi$ having a nearly-normal distribution when sampled at random times, as shown in Fig. \ref{fig:xi-dist-numeric}, with $\Psi_i\lesssim m_\phi\lesssim \Phi$.
% Figure \ref{fig:xi-dist-numeric} shows an example of the randomness of $\xi$.
Although $\xi$ is a stochastic variable, using Eqn. (\ref{eqn:ctc}) we can say that $\xi_n$ is constant for each oscillation of $\phi$ but varies between periods $n$, and the resulting inflaton solution is
\begin{align}
    \phi_n(t) = \Phi \cos\left(m_\phi t \sqrt{1+\xi_n}\right)
\end{align}
and the average difference from the noiseless case over one period $T$
%, with $t\in(0, 2\pi/m_\phi)$,
is
\begin{align}
    \frac{\langle\delta\phi\rangle_T}{\Phi} = \left\langle \cos\left(m_\phi t\sqrt{1+\xi}\right) - \cos(m_\phi t)\right\rangle_T = \frac{1}{2}\xi + \mathcal{O}\left(\xi^2\right)\ .
\end{align}
The EOM of the reheat field $\chi$ is
\begin{align}
     \frac{\dd^2\chi_k}{\dd t^2}+ \left[ k^2+m_\chi^2 + 2\sigma \Phi \cos( m t\sqrt{1+\xi_n})\right] \chi_k = 0\ .
\end{align}
If we define $2\tau=m_\phi t\sqrt{1+\xi_n}$, then we can rewrite the EOM as Mathieu's equation,
\begin{align}
    \frac{\dd^2\chi_k}{\dd \tau^2} + [A_n+2q_n\cos(2\tau)] \chi_k=0
    \qquad \text{where} \qquad 
    A_n = \frac{4(k^2+m_\chi^2)}{m_\phi^2(1+\xi_n)} \ \ ,
    \quad   q_n=\frac{4\sigma\Phi}{m_\phi^2(1+\xi_n)}\ ,
\end{align}
just as in Eqn. (\ref{eqn:hills}), in the $\lambda=0$ case, but with a modified mass;
now we have the relative addition $\xi$, which collects the effect of the fields $\psi_i$ on the inflaton solution $\phi$ and is manifest as a cycle-to-cycle variation in the Mathieu's equation parameters $A$ and $q$.
We can see from Fig. \ref{fig:xi-dist-numeric} that we can approximately treat $\xi$ as a normally-distributed parameter that is independently drawn each cycle.
The results in the following section use this assumption with a conservative $\sqrt{\text{Var}\ \xi}=0.05$.

\section{Numerical Results} \label{sec:results}

This section shows slices of Floquet charts like Fig. \ref{fig:quartplane} using a normal distribution of $\xi$ with a mean of zero and a standard deviation of $\sqrt{\text{Var}\ \xi}=0.05$. 
These slices help us to see more accurately the effects of fluctuations on the Floquet exponents, although the maps themselves remain qualitatively useful.

We see that regions of stability, both narrow and broad, obtain nonzero growth rates to ensure that reheating is able to start and convert more inflaton energy into reheat field energy; these growth rates are dominated by the random walk term $\gamma_R$ in Eqn. (\ref{eqn:gamma-parts}), allowing for a nonzero growth rate despite the local unfluctuated average growth rate of zero.
Regions of broad stability experience decreased growth rates with fluctuations because their new growth rate is dominated by the $\gamma_\infty$ term in Eqn. (\ref{eqn:gamma-parts}).
These asymptotic (local average) growth rates get contributions from nearby regions on the Floquet map with lower growth rate.
However, the resonance effect is still strong in the broad resonance regimes.

\section{Summary and Discussion}
\label{sec:conclusion}

This paper explores how stochasticity in Hill's equation can, across a wide range of parameter space, influence the efficiency of inflationary (p)reheating. This approach addresses some of the model dependence that arises when exploring how (p)reheating occurs and how efficiently it operates at the end of inflation. In particular, in the presence of fluctuations, successful reheating is more likely to occur for a greater range of parameters, such as those  motivated by UV complete models of inflation. 

This approach can also inform us about the nature of the cosmological moduli problem and affect when the universe thermalized, leading to the radiation dominated phase needed for Big Bang Nucleosythesis.  We also found  that non-perturbative particle production could benefit from stochasticity, which increases particle production rates in otherwise nonviable regions of parameter spaces, in the context of both (p)reheating and the cosmological moduli problem. In the case of the former,  the possibility of long-lasting early matter domination disagrees with CMB observations and endangers UV complete scenarios for the early universe \cite{COUGHLAN198359}.

Parametric resonance is a possible non-perturbative solution to the cosmological moduli problem \cite{2017-gauge-CMP, deal2025cosmologicalmodulinonperturbativeproduction} that could benefit from additional growth from fluctuations since the preferred parameter space is broadly stable.
More generally, Hill's equation is ubiquitous in physical systems with stochasticity, and we wish to emphasize the importance of considering fluctuations in periodic systems.

Although our results apply more generally to any source of fluctuations that modify Hill's equation, our specific approach utilized coupling to multiple scalar fields. The consideration of such fields is well motivated by UV complete theories, as well as approaches to addressing important issues such as the Electroweak Hierarchy problem \cite{Dimopoulos:2005ac,Arkani-Hamed:2016rle}. We also note that another motivation was to build upon previous work, which has been primarily based on numerical treatments  \cite{Gu:2018anb,Dasgupta:2004dw,Allahverdi:2004ds,Allahverdi:2010xz,Braden:2010wd,Dufaux_2006}. The results of this paper suggest that including stochastic fluctuations within the Hill's equation can further our understanding of the transition from Inflation to the Hot Big Bang.

\appendix
\section{Stochastic Differential Equations and Cycle to Cycle Variations}

In this appendix we show how continuously varying noise is equivalent to cycle-to-cycle variation in an underlying periodic system, allowing us to bypass the computation of stochastic integrals.

The general Hill's equation with a periodic effective frequency $\Omega_n^2$ that varies between periods $n$, all with the same period, is
\begin{align}
    \frac{\text{d}^2y}{\text{d}t^2} + \Omega_n^2(t) y(t) = \xi
\end{align}
We can write the periodic coefficient as $\Omega^2_n(t)=A_n+2q_nP(t)$, where $P(t)$ has periodicity $T$.
The parameters $A_n=A+\ell_n$ and $q_n=q+p_n$ can vary from cycle to cycle with means $\langle l_k\rangle=\langle p_k\rangle=0$.
If we instead think of Hill's equation with constant parameters and a single stochastic term $\xi$ with $\langle\xi\rangle=0$,
\begin{align}
    \frac{\text{d}^2y}{\text{d}t^2} + \left[A+2q P(t)\right] y(t) = \xi \ \ ,
\end{align}
we could relate this stochastic differential equation to the cycle-to-cycle parameters $\ell_n$ and $p_n$ like~\cite{Adams_2013}
\begin{align}
    \ell_n=\frac{J_2\Xi_1-J_1\Xi_2}{I_1J_2-I_2J_1}\ \ ,\quad 
    p_n=\frac{I_1\Xi_2-I_2\Xi_1}{I_1J_2-I_2J_1}
    \label{eqn:ctc}
\end{align}
where
\begin{align}
    I_i=\int_0^T u_i^2(t)\text{d}t\ \ ,\quad
    J_i=\int_0^T P(t) u_i^2(t)\text{d}t\ \ ,\quad
    \Xi_i=\int_0^T \xi u_i(t)\text{d}t\ \ .
\end{align}
These stochastic integrals may be evaluated by methods attributed to Ito and Stratonovich. 
However, the existence of this correspondence to cycle-to-cycle variations allows us to avoid the comparatively arduous evaluation of stochastic integrals and proceed with the cycle-to-cycle analysis presented in this work.

\section{Growth Rates from Fluctuations in the Classically Stable Regime} \label{sec:stable-flucs}

This appendix shows how fluctuations can result in nonzero growth rates for Hill's equation in regimes of parameter space that would be stable in their absence. Recall that the total growth rate of a fluctuating system is composed of an average over individual cycles $\gamma_\infty$ and a random walk term $\gamma_R$. In the stable regime, the matrix elements $|h_n|<1$, since $\gamma_\infty=0$. The growth rate is thus dominated by $\gamma_R$, which we will find in this section in the limit of small but nonzero fluctuations.

To start, since $|h_n|<1$, we can define $h_n=\cos\theta_n$ and $g_nL_n=\sin\theta_n$ (see \cite{AdBloch2009}), so that the transfer matrix can be written in the form 
\begin{align}
    \mathbb{M}_n = \begin{bmatrix}
        \cos\theta_n & -L_n\sin\theta_n \\ \sin\theta_n/L_n &\cos\theta_n
    \end{bmatrix}\,.
    \label{eqn:cosM}
\end{align}
Each matrix thus has the form of an elliptical rotation matrix with a length parameter $L_n$ and an angle $\theta_n$ that vary between cycles.

All of our closely related stochastic variables, $x=A,q,h,g,\theta,...$, with means $x_0\equiv\langle x\rangle$, have deviations between cycles labeled $n$, $x_n=(1+\delta_{xn})x_0$. All of the fluctuating parts $\delta_x$ are of similar magnitude to each other, and all have zero mean $\langle \delta_x\rangle=0$. For our purposes, all $|\delta_{xn}|\ll1$ and we will find the leading order correction to the growth rate. Of particular interest in our case is the variation in the length parameter $L_n$, which can be written in the form  
\begin{align}
    L_n\equiv (1+\delta_{Ln})L_0 \qquad \text{where}\qquad L_0=\left\langle \frac{\sin\theta}{g} \right\rangle
\end{align}
This ansatz allows us to split the transfer $\mathbb{M}_n$ into an a true elliptical rotation matrix $\mathbb{E}_n$, which has constant length parameter $L_0$, and correction terms in powers of $\delta_{L}$, denoted by the superscripts in the second line below,
\begin{align}
    \mathbb{M}_n &=
    \begin{bmatrix}
        \cos\theta_n & -L_0\sin\theta_n \\ \sin\theta_n/L_0 & \cos\theta_n
    \end{bmatrix}
    -\delta_{Ln} \sin\theta_n
    \begin{bmatrix}
        0 & L_0 \\ 1/L_0 & 0
    \end{bmatrix}
     + \delta_{Ln}^2 \sin\theta_n 
    \begin{bmatrix}
        0 & 0 \\ 1/L_0 & 0
    \end{bmatrix}
    +\mathcal{O}\left(\delta^3\right) \nonumber \\
    &\equiv \mathbb{E}_n + \mathbb{M}_n^1 + \mathbb{M}_n^2 + \mathcal{O}\left(\delta^3\right) \ .
\end{align}
After the elapse of $N$ cycles, the total transfer matrix is
\begin{align}
    \mathbb{M}_{(N)} \equiv \prod_{n=1}^N \mathbb{M}_n = \prod_{n=1}^N\mathbb{E}_n + \sum_{n=1}^N \mathbb{P}_n + \sum_{n,k=1}^N \mathbb{Q}_{nk} + \mathcal{O}(\delta^3)\ .
    \label{eqn:M(N)}
\end{align}
The parenthetical subscript defined by the first equality is distinguishes $\mathbb{M}_{(N)}$ from $\mathbb{M}_N$, the $N$th matrix in the product.
We define $\mathbb{P}_n$ and $\mathbb{Q}_{nk}$ as
\begin{align}
    \mathbb{P}_n = 
    \left(\prod_{\ell=n+1}^{N} \mathbb{E}_\ell\right) \left(\mathbb{M}_n^1+\mathbb{M}_n^2\right) 
    \left(\prod_{\ell=1}^{n-1} \mathbb{E}_\ell\right)
\end{align}
and
\begin{align}
    \mathbb{Q}_{nk} = 
    \left(\prod_{\ell=n+1}^{N} \mathbb{E}_\ell\right) 
    \left(\mathbb{M}_n^1+\mathbb{M}_n^2\right)
    \left(\prod_{\ell=k+1}^{n-1} \mathbb{E}_\ell\right)
    \left(\mathbb{M}_k^1+\mathbb{M}_k^2\right)
    \left(\prod_{\ell=1}^{k-1} \mathbb{E}_\ell\right)\ .
\end{align}
Using the following notation for a sum of angles,
\begin{align}
    \Theta_k = \sum_{\ell=1}^k \theta_\ell
\end{align}
we get the familiar result for the product of elliptical matrices,
\begin{align}
    \prod_{n=1}^N\mathbb{E}_n = 
    \begin{bmatrix}
        \cos\Theta_N & -L_0\sin\Theta_N \\ \sin\Theta_N/L_0 & \cos\Theta_N
    \end{bmatrix}\ .
    \label{eqn:EN}
\end{align}
Since $\langle\delta_L\rangle=0$, any sums in Eqn. (\ref{eqn:M(N)}) over terms linear in $\delta_L$ vanish, so that
\begin{align}
    \sum_{n=1}^N\mathbb{P}_n = 
    \sum_{n=1}^N\left(\prod_{\ell=n+1}^{N} \mathbb{E}_\ell\right)  
    \mathbb{M}_n^2 
    \left(\prod_{\ell=1}^{n-1} \mathbb{E}_\ell\right)
    \quad \text{and} \quad
    \sum_{n,k=1}^N\mathbb{Q}_{nk} = 0 + \mathcal{O}\left(\delta^4\right)\ 
\end{align}
so that the lowest leading correcting order becomes $\delta_L^2$. Using this expression for $\sum_n\mathbb{P}_n$ and the product of elliptical matrices in Eqn. (\ref{eqn:EN}), we get
\begin{align}
    \sum_{n=1}^N\mathbb{P}_n = \sum_{n=1}^N \delta_{Ln}^2 \sin\theta_n
    \begin{bmatrix}
        -\sin(\Theta_N-\Theta_n)\cos\Theta_{n-1}
        && L_0 \sin(\Theta_N-\Theta_n)\sin\Theta_{n-1} \\
        \cos(\Theta_N-\Theta_n)\cos\Theta_{n-1}/L_0
        && -\cos(\Theta_N-\Theta_n)\sin\Theta_{n-1}
    \end{bmatrix}
\end{align}
The resulting eigenvalue of $\mathbb{M}_{(N)}$ to $\mathcal{O}(\delta^2)$, the lowest leading order, is then given by 
\begin{align}
    \lambda_{(N)} = \cos\Theta_N \pm i\sin\Theta_N + \frac{1}{2} \sum_n \delta_{Ln}^2\sin\theta_n \big[-\sin(\Theta_N-\theta_n) \pm i\cos(\Theta_N-\theta_n) \big]
\end{align}
with magnitude
\begin{align}
    \left|\lambda_{(N)}\right|^2 = 1 \pm \sum_{n=1}^N\delta_{Ln}^2 \sin^2 \theta_n
    = 1\pm N \left\langle \delta_L^2\sin^2\theta \right\rangle\ ,
\end{align}
where the second equality holds in the limit of infinite $N$, which ensures that we sample the entire space of $\delta_L$.
The effective growth rate per cycle can then be written as 
\begin{align}
    e^{\gamma_R T} \equiv \left|\lambda_{(N)}\right|^{1/N} = 1 + \frac{1}{2} \left\langle \delta_L^2\sin^2\theta \right\rangle
\end{align}
which to leading order matches the definitions of growth rate given in Eqn. (\ref{eqn:gama-loglam}) and Eqn. (\ref{eqn:gama-normM}), i.e., 
\begin{align}
    \gamma_R = \frac{1}{NT} \lim_{N\to\infty} \log \left|\lambda_{(N)}\right|
    = \frac{1}{2T} \left\langle \delta_L^2\sin^2\theta \right\rangle\ .
    \label{eqn:gammaR}
\end{align}
Fittingly, this random walk growth rate depends on the variance of fluctuations and is a positive definite addition to the total growth rate. Moreover, in the regime where the equation is otherwise stable, this contribution provides a nonzero growth rate.  

\begin{figure}
    \centering
    \includegraphics[width=0.5\linewidth]{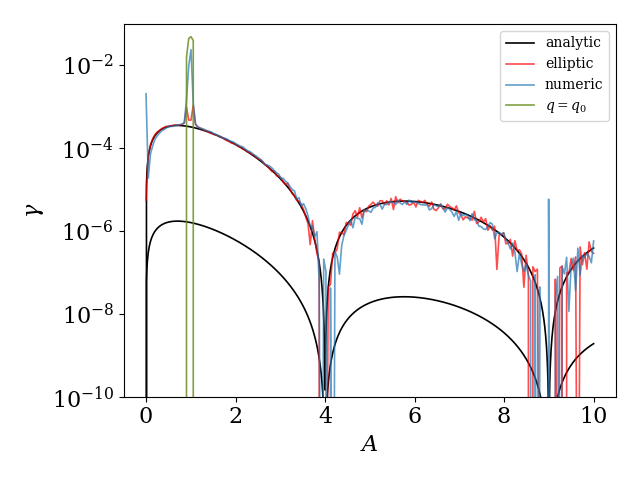}
    \caption{Growth rates vs. nonstochastic $A=A_0$ for $q_0=0.007,\ 0.1$ and a uniform selection of $q\in(0,q_0)$ using Eqn. (\ref{eqn:ellpt-gammaR}) (black), multiplying random matrices of the form Eqn. (\ref{eqn:cosM}) using Eqn. (\ref{eqn:ellpt-L}) for $q_0=0.1$, (red), multiplying random matrices of the form of Eqn. (\ref{eqn:transfer}) found by numerically solving one period of Mathieu's equation for $q_0=0.1$ (blue), and the unfluctuated growth rate for $q=0.1$ (green). Despite some numeric noise, we see that Eqn. (\ref{eqn:ellpt-gammaR}) is a successful approximation of the random walk growth rate while acknowledging some features that appear in the numeric solution at square integer values of $A$, especially $A=1$ around which the unfluctuated growth rate is nonzero.}
    \label{fig:ellpt-numeric}
\end{figure}

As a concrete example, let us calculate the random walk growth rate for Mathieu's equation in the limit that $q,q_0\ll1$ with constant $A=\omega_0^2$,
\begin{align}
    \ddot u(t) +\left[\omega_0^2+2q\cos(2t)\right]u(t)=0\ .
\end{align}
We'll expand the solution in orders of $q$, with initial conditions $u(0)=1$ and $\dot u(0)=0$
\begin{align}
    u(t)=u_{(0)}(t)+u_{(1)}(t)+\mathcal{O}(q^2) \quad \text{where} \quad u_{(0)}(t)=\alpha\cos(\omega_0 t)
\end{align}
and the differential equation of interest becomes 
\begin{align}
    \ddot u_{(1)}+\omega_0^2u_{(1)} +2q\alpha\cos(2t)\cos(\omega_0t)+\mathcal{O}(q^2)=0\ .
\end{align}
Then the solution with the desired initial condition, to first order in $q$, is
\begin{align}
    u(t) = \frac{1}{2(\omega_0^2-1)-q}\left(2(\omega_0^2-1)\cos(\omega_0t) - q [\cos(2t)\cos(\omega_0t)+\omega\sin(2t)\sin(\omega_0t)]\right)\ .
\end{align}
Then the transfer matrix takes the form of Eqn. (\ref{eqn:cosM}) with constant elliptic rotation angle $\theta=\omega_0\pi$ and shape parameter
\begin{align}
    L=\omega_0\frac{2(\omega_0^2-1)+q}{2(\omega_0^2-1)-q}
    \label{eqn:ellpt-L}
\end{align}
that varies from cycle to cycle with varying $q$.
Then $\delta_{L_n}$ is 
\begin{align}
    \delta_{L_n} = \frac{q-q_0}{\omega_0^2-1}\ .
\end{align}
Using Eqn. (\ref{eqn:gammaR}), we get
\begin{align}
    \gamma_R=\frac{1}{2\pi}\frac{\text{Var }q}{(\omega_0^2-1)^2}\sin^2(\omega_0\pi)\ ,
    \label{eqn:ellpt-gammaR}
\end{align}
the growth rate resulting from fluctuations where $\gamma_0=0$ and $q\ll1$, verified in Fig. \ref{fig:ellpt-numeric}.

\section*{Acknowledgements}
We thank Tom Giblin for useful discussions. S.W. thanks the Simons Center and the University of South Carolina for hospitality. This research was supported in part by DOE grant DE-FG02-85ER40237. A.B. was partially supported by NSF grant  DMS-2103026, and AFOSR grants FA
9550-22-1-0215 and FA 9550-23-1-0400. L.B. and F.C.A. were supported in part by the Leinweber Institute for Theoretical Physics at the University of Michigan.

\bibliography{shortref.bib}

\begin{thebibliography}{46}
\expandafter\ifx\csname natexlab\endcsname\relax\def\natexlab#1{#1}\fi
\expandafter\ifx\csname bibnamefont\endcsname\relax
  \def\bibnamefont#1{#1}\fi
\expandafter\ifx\csname bibfnamefont\endcsname\relax
  \def\bibfnamefont#1{#1}\fi
\expandafter\ifx\csname citenamefont\endcsname\relax
  \def\citenamefont#1{#1}\fi
\expandafter\ifx\csname url\endcsname\relax
  \def\url#1{\texttt{#1}}\fi
\expandafter\ifx\csname urlprefix\endcsname\relax\def\urlprefix{URL }\fi
\providecommand{\bibinfo}[2]{#2}
\providecommand{\eprint}[2][]{\url{#2}}

\bibitem[{\citenamefont{{Guth}}(1981)}]{Guth1981}
\bibinfo{author}{\bibfnamefont{A.~H.} \bibnamefont{{Guth}}},
  \bibinfo{journal}{\prd} \textbf{\bibinfo{volume}{23}}, \bibinfo{pages}{347}
  (\bibinfo{year}{1981}).

\bibitem[{\citenamefont{Bardeen et~al.}(1983)\citenamefont{Bardeen, Steinhardt,
  and Turner}}]{Bardeen83}
\bibinfo{author}{\bibfnamefont{J.~M.} \bibnamefont{Bardeen}},
  \bibinfo{author}{\bibfnamefont{P.~J.} \bibnamefont{Steinhardt}},
  \bibnamefont{and} \bibinfo{author}{\bibfnamefont{M.~S.}
  \bibnamefont{Turner}}, \bibinfo{journal}{Phys. Rev. D}
  \textbf{\bibinfo{volume}{28}}, \bibinfo{pages}{679} (\bibinfo{year}{1983}),
  \urlprefix\url{https://link.aps.org/doi/10.1103/PhysRevD.28.679}.

\bibitem[{\citenamefont{{Planck Collaboration}}(2020)}]{plank2018inflation}
\bibinfo{author}{\bibnamefont{{Planck Collaboration}}}, \bibinfo{journal}{A\&A}
  \textbf{\bibinfo{volume}{641}}, \bibinfo{pages}{A10} (\bibinfo{year}{2020}),
  \urlprefix\url{https://doi.org/10.1051/0004-6361/201833887}.

\bibitem[{\citenamefont{Albrecht and Steinhardt}(1982)}]{AS_newinfl}
\bibinfo{author}{\bibfnamefont{A.}~\bibnamefont{Albrecht}} \bibnamefont{and}
  \bibinfo{author}{\bibfnamefont{P.~J.} \bibnamefont{Steinhardt}},
  \bibinfo{journal}{Phys. Rev. Lett.} \textbf{\bibinfo{volume}{48}},
  \bibinfo{pages}{1220} (\bibinfo{year}{1982}),
  \urlprefix\url{https://link.aps.org/doi/10.1103/PhysRevLett.48.1220}.

\bibitem[{\citenamefont{{Linde}}(1982)}]{Linde_newinfl}
\bibinfo{author}{\bibfnamefont{A.~D.} \bibnamefont{{Linde}}},
  \bibinfo{journal}{Physics Letters B} \textbf{\bibinfo{volume}{108}},
  \bibinfo{pages}{389} (\bibinfo{year}{1982}).

\bibitem[{\citenamefont{Traschen and Brandenberger}(1990)}]{Traschen:1990sw}
\bibinfo{author}{\bibfnamefont{J.~H.} \bibnamefont{Traschen}} \bibnamefont{and}
  \bibinfo{author}{\bibfnamefont{R.~H.} \bibnamefont{Brandenberger}},
  \bibinfo{journal}{Phys. Rev. D} \textbf{\bibinfo{volume}{42}},
  \bibinfo{pages}{2491} (\bibinfo{year}{1990}).

\bibitem[{\citenamefont{Kofman et~al.}(1994)\citenamefont{Kofman, Linde, and
  Starobinsky}}]{Kofman94}
\bibinfo{author}{\bibfnamefont{L.}~\bibnamefont{Kofman}},
  \bibinfo{author}{\bibfnamefont{A.}~\bibnamefont{Linde}}, \bibnamefont{and}
  \bibinfo{author}{\bibfnamefont{A.~A.} \bibnamefont{Starobinsky}},
  \bibinfo{journal}{Phys. Rev. Lett.} \textbf{\bibinfo{volume}{73}},
  \bibinfo{pages}{3195} (\bibinfo{year}{1994}),
  \urlprefix\url{https://link.aps.org/doi/10.1103/PhysRevLett.73.3195}.

\bibitem[{\citenamefont{Kofman et~al.}(1997)\citenamefont{Kofman, Linde, and
  Starobinsky}}]{Kofman:1997yn}
\bibinfo{author}{\bibfnamefont{L.}~\bibnamefont{Kofman}},
  \bibinfo{author}{\bibfnamefont{A.~D.} \bibnamefont{Linde}}, \bibnamefont{and}
  \bibinfo{author}{\bibfnamefont{A.~A.} \bibnamefont{Starobinsky}},
  \bibinfo{journal}{Phys. Rev. D} \textbf{\bibinfo{volume}{56}},
  \bibinfo{pages}{3258} (\bibinfo{year}{1997}), \eprint{hep-ph/9704452}.

\bibitem[{\citenamefont{{Zanchin} et~al.}(1998)\citenamefont{{Zanchin}, {Maia},
  {Craig}, and {Brandenberger}}}]{Zanchin1998}
\bibinfo{author}{\bibfnamefont{V.}~\bibnamefont{{Zanchin}}},
  \bibinfo{author}{\bibfnamefont{J.}~\bibnamefont{{Maia}}, \bibfnamefont{A.}},
  \bibinfo{author}{\bibfnamefont{W.}~\bibnamefont{{Craig}}}, \bibnamefont{and}
  \bibinfo{author}{\bibfnamefont{R.}~\bibnamefont{{Brandenberger}}},
  \bibinfo{journal}{\prd} \textbf{\bibinfo{volume}{57}}, \bibinfo{pages}{4651}
  (\bibinfo{year}{1998}), \eprint{hep-ph/9709273}.

\bibitem[{\citenamefont{Zanchin et~al.}(1999)\citenamefont{Zanchin, Maia,
  Craig, and Brandenberger}}]{Zanchin_1999}
\bibinfo{author}{\bibfnamefont{V.}~\bibnamefont{Zanchin}},
  \bibinfo{author}{\bibfnamefont{A.}~\bibnamefont{Maia}},
  \bibinfo{author}{\bibfnamefont{W.}~\bibnamefont{Craig}}, \bibnamefont{and}
  \bibinfo{author}{\bibfnamefont{R.}~\bibnamefont{Brandenberger}},
  \bibinfo{journal}{Physical Review D} \textbf{\bibinfo{volume}{60}}
  (\bibinfo{year}{1999}), ISSN \bibinfo{issn}{1089-4918},
  \urlprefix\url{http://dx.doi.org/10.1103/PhysRevD.60.023505}.

\bibitem[{\citenamefont{Furstenberg and
  Kesten}(1960)}]{furstenberg_random_matrices}
\bibinfo{author}{\bibfnamefont{H.}~\bibnamefont{Furstenberg}} \bibnamefont{and}
  \bibinfo{author}{\bibfnamefont{H.}~\bibnamefont{Kesten}},
  \bibinfo{journal}{The Annals of Mathematical Statistics}
  \textbf{\bibinfo{volume}{31}}, \bibinfo{pages}{457 } (\bibinfo{year}{1960}),
  \urlprefix\url{https://doi.org/10.1214/aoms/1177705909}.

\bibitem[{\citenamefont{{Adams} and {Bloch}}(2008)}]{AdBloch2008}
\bibinfo{author}{\bibfnamefont{F.~C.} \bibnamefont{{Adams}}} \bibnamefont{and}
  \bibinfo{author}{\bibfnamefont{A.~M.} \bibnamefont{{Bloch}}},
  \bibinfo{journal}{SIAM J. Appl. Math.} \textbf{\bibinfo{volume}{69}},
  \bibinfo{pages}{947} (\bibinfo{year}{2008}).

\bibitem[{\citenamefont{Murakami and
  Terao}(2016)}]{murakami2016nonperturbativeevaluationquantumparticle}
\bibinfo{author}{\bibfnamefont{A.}~\bibnamefont{Murakami}} \bibnamefont{and}
  \bibinfo{author}{\bibfnamefont{H.}~\bibnamefont{Terao}},
  \emph{\bibinfo{title}{Nonperturbative evaluation of quantum particle
  production in parametric resonance enhanced by noise}}
  (\bibinfo{year}{2016}), \eprint{1604.01886},
  \urlprefix\url{https://arxiv.org/abs/1604.01886}.

\bibitem[{\citenamefont{Anderson}(1958)}]{anderson_loc}
\bibinfo{author}{\bibfnamefont{P.~W.} \bibnamefont{Anderson}},
  \bibinfo{journal}{Phys. Rev.} \textbf{\bibinfo{volume}{109}},
  \bibinfo{pages}{1492} (\bibinfo{year}{1958}),
  \urlprefix\url{https://link.aps.org/doi/10.1103/PhysRev.109.1492}.

\bibitem[{\citenamefont{Amin and Baumann}(2015)}]{Amin2015FromWT}
\bibinfo{author}{\bibfnamefont{M.}~\bibnamefont{Amin}} \bibnamefont{and}
  \bibinfo{author}{\bibfnamefont{D.}~\bibnamefont{Baumann}},
  \bibinfo{journal}{Journal of Cosmology and Astroparticle Physics}
  \textbf{\bibinfo{volume}{2016}}, \bibinfo{pages}{045 }
  (\bibinfo{year}{2015}),
  \urlprefix\url{https://api.semanticscholar.org/CorpusID:85441022}.

\bibitem[{\citenamefont{Amin et~al.}(2017)\citenamefont{Amin, Garcia, Xie, and
  Wen}}]{Amin2017MultifieldSP}
\bibinfo{author}{\bibfnamefont{M.}~\bibnamefont{Amin}},
  \bibinfo{author}{\bibfnamefont{M.~A.} \bibnamefont{Garcia}},
  \bibinfo{author}{\bibfnamefont{H.-Y.} \bibnamefont{Xie}}, \bibnamefont{and}
  \bibinfo{author}{\bibfnamefont{O.}~\bibnamefont{Wen}},
  \bibinfo{journal}{Journal of Cosmology and Astroparticle Physics}
  \textbf{\bibinfo{volume}{2017}}, \bibinfo{pages}{015 }
  (\bibinfo{year}{2017}),
  \urlprefix\url{https://api.semanticscholar.org/CorpusID:118960064}.

\bibitem[{\citenamefont{Gu and Brandenberger}(2020)}]{Gu_2020}
\bibinfo{author}{\bibfnamefont{B.-M.} \bibnamefont{Gu}} \bibnamefont{and}
  \bibinfo{author}{\bibfnamefont{R.}~\bibnamefont{Brandenberger}},
  \bibinfo{journal}{Chinese Physics C} \textbf{\bibinfo{volume}{44}},
  \bibinfo{pages}{015103} (\bibinfo{year}{2020}),
  \urlprefix\url{https://dx.doi.org/10.1088/1674-1137/44/1/015103}.

\bibitem[{\citenamefont{Easther and McAllister}(2006)}]{Easther:2005zr}
\bibinfo{author}{\bibfnamefont{R.}~\bibnamefont{Easther}} \bibnamefont{and}
  \bibinfo{author}{\bibfnamefont{L.}~\bibnamefont{McAllister}},
  \bibinfo{journal}{JCAP} \textbf{\bibinfo{volume}{05}}, \bibinfo{pages}{018}
  (\bibinfo{year}{2006}), \eprint{hep-th/0512102}.

\bibitem[{\citenamefont{Arkani-Hamed et~al.}(2016)\citenamefont{Arkani-Hamed,
  Cohen, D'Agnolo, Hook, Kim, and Pinner}}]{Arkani-Hamed:2016rle}
\bibinfo{author}{\bibfnamefont{N.}~\bibnamefont{Arkani-Hamed}},
  \bibinfo{author}{\bibfnamefont{T.}~\bibnamefont{Cohen}},
  \bibinfo{author}{\bibfnamefont{R.~T.} \bibnamefont{D'Agnolo}},
  \bibinfo{author}{\bibfnamefont{A.}~\bibnamefont{Hook}},
  \bibinfo{author}{\bibfnamefont{H.~D.} \bibnamefont{Kim}}, \bibnamefont{and}
  \bibinfo{author}{\bibfnamefont{D.}~\bibnamefont{Pinner}},
  \bibinfo{journal}{Phys. Rev. Lett.} \textbf{\bibinfo{volume}{117}},
  \bibinfo{pages}{251801} (\bibinfo{year}{2016}), \eprint{1607.06821}.

\bibitem[{\citenamefont{Weinberg}(2008)}]{weinberg_EFTinfl}
\bibinfo{author}{\bibfnamefont{S.}~\bibnamefont{Weinberg}},
  \bibinfo{journal}{Phys. Rev. D} \textbf{\bibinfo{volume}{77}},
  \bibinfo{pages}{123541} (\bibinfo{year}{2008}),
  \urlprefix\url{https://link.aps.org/doi/10.1103/PhysRevD.77.123541}.

\bibitem[{\citenamefont{Cheung et~al.}(2008)\citenamefont{Cheung, Creminelli,
  Fitzpatrick, Kaplan, and Senatore}}]{Cheung:2007st}
\bibinfo{author}{\bibfnamefont{C.}~\bibnamefont{Cheung}},
  \bibinfo{author}{\bibfnamefont{P.}~\bibnamefont{Creminelli}},
  \bibinfo{author}{\bibfnamefont{A.~L.} \bibnamefont{Fitzpatrick}},
  \bibinfo{author}{\bibfnamefont{J.}~\bibnamefont{Kaplan}}, \bibnamefont{and}
  \bibinfo{author}{\bibfnamefont{L.}~\bibnamefont{Senatore}},
  \bibinfo{journal}{JHEP} \textbf{\bibinfo{volume}{03}}, \bibinfo{pages}{014}
  (\bibinfo{year}{2008}), \eprint{0709.0293}.

\bibitem[{\citenamefont{\"Ozsoy et~al.}(2015)\citenamefont{\"Ozsoy, Sinha, and
  Watson}}]{Ozsoy:2014sba}
\bibinfo{author}{\bibfnamefont{O.}~\bibnamefont{\"Ozsoy}},
  \bibinfo{author}{\bibfnamefont{K.}~\bibnamefont{Sinha}}, \bibnamefont{and}
  \bibinfo{author}{\bibfnamefont{S.}~\bibnamefont{Watson}},
  \bibinfo{journal}{Phys. Rev. D} \textbf{\bibinfo{volume}{91}},
  \bibinfo{pages}{103509} (\bibinfo{year}{2015}), \eprint{1410.0016}.

\bibitem[{\citenamefont{\"Ozsoy et~al.}(2017)\citenamefont{\"Ozsoy, Giblin,
  Nesbit, \c{S}eng\"or, and Watson}}]{Ozsoy:2017mqc}
\bibinfo{author}{\bibfnamefont{O.}~\bibnamefont{\"Ozsoy}},
  \bibinfo{author}{\bibfnamefont{J.~T.} \bibnamefont{Giblin}},
  \bibinfo{author}{\bibfnamefont{E.}~\bibnamefont{Nesbit}},
  \bibinfo{author}{\bibfnamefont{G.}~\bibnamefont{\c{S}eng\"or}},
  \bibnamefont{and} \bibinfo{author}{\bibfnamefont{S.}~\bibnamefont{Watson}},
  \bibinfo{journal}{Phys. Rev. D} \textbf{\bibinfo{volume}{96}},
  \bibinfo{pages}{123524} (\bibinfo{year}{2017}), \eprint{1701.01455}.

\bibitem[{\citenamefont{Lopez~Nacir et~al.}(2012)\citenamefont{Lopez~Nacir,
  Porto, Senatore, and Zaldarriaga}}]{LopezNacir:2011kk}
\bibinfo{author}{\bibfnamefont{D.}~\bibnamefont{Lopez~Nacir}},
  \bibinfo{author}{\bibfnamefont{R.~A.} \bibnamefont{Porto}},
  \bibinfo{author}{\bibfnamefont{L.}~\bibnamefont{Senatore}}, \bibnamefont{and}
  \bibinfo{author}{\bibfnamefont{M.}~\bibnamefont{Zaldarriaga}},
  \bibinfo{journal}{JHEP} \textbf{\bibinfo{volume}{01}}, \bibinfo{pages}{075}
  (\bibinfo{year}{2012}), \eprint{1109.4192}.

\bibitem[{\citenamefont{Kane et~al.}(2015)\citenamefont{Kane, Sinha, and
  Watson}}]{Kane:2015jia}
\bibinfo{author}{\bibfnamefont{G.}~\bibnamefont{Kane}},
  \bibinfo{author}{\bibfnamefont{K.}~\bibnamefont{Sinha}}, \bibnamefont{and}
  \bibinfo{author}{\bibfnamefont{S.}~\bibnamefont{Watson}},
  \bibinfo{journal}{Int. J. Mod. Phys. D} \textbf{\bibinfo{volume}{24}},
  \bibinfo{pages}{1530022} (\bibinfo{year}{2015}), \eprint{1502.07746}.

\bibitem[{\citenamefont{Giblin et~al.}(2017)\citenamefont{Giblin, Kane, Nesbit,
  Watson, and Zhao}}]{2017-gauge-CMP}
\bibinfo{author}{\bibfnamefont{J.~T.} \bibnamefont{Giblin}},
  \bibinfo{author}{\bibfnamefont{G.}~\bibnamefont{Kane}},
  \bibinfo{author}{\bibfnamefont{E.}~\bibnamefont{Nesbit}},
  \bibinfo{author}{\bibfnamefont{S.}~\bibnamefont{Watson}}, \bibnamefont{and}
  \bibinfo{author}{\bibfnamefont{Y.}~\bibnamefont{Zhao}},
  \bibinfo{journal}{Phys. Rev. D} \textbf{\bibinfo{volume}{96}},
  \bibinfo{pages}{043525} (\bibinfo{year}{2017}),
  \urlprefix\url{https://link.aps.org/doi/10.1103/PhysRevD.96.043525}.

\bibitem[{\citenamefont{Deal et~al.}(2025)\citenamefont{Deal, Barrowes, John
  T.~Giblin, Sinha, Watson, and
  Adams}}]{deal2025cosmologicalmodulinonperturbativeproduction}
\bibinfo{author}{\bibfnamefont{R.~W.} \bibnamefont{Deal}},
  \bibinfo{author}{\bibfnamefont{L.}~\bibnamefont{Barrowes}},
  \bibinfo{author}{\bibfnamefont{J.}~\bibnamefont{John T.~Giblin}},
  \bibinfo{author}{\bibfnamefont{K.}~\bibnamefont{Sinha}},
  \bibinfo{author}{\bibfnamefont{S.}~\bibnamefont{Watson}}, \bibnamefont{and}
  \bibinfo{author}{\bibfnamefont{F.~C.} \bibnamefont{Adams}},
  \emph{\bibinfo{title}{Cosmological moduli and non-perturbative production of
  axions}} (\bibinfo{year}{2025}), \eprint{2501.17229},
  \urlprefix\url{https://arxiv.org/abs/2501.17229}.

\bibitem[{\citenamefont{Shuhmaher and Brandenberger}(2006)}]{Shuhmaher_2006}
\bibinfo{author}{\bibfnamefont{N.}~\bibnamefont{Shuhmaher}} \bibnamefont{and}
  \bibinfo{author}{\bibfnamefont{R.}~\bibnamefont{Brandenberger}},
  \bibinfo{journal}{Physical Review D} \textbf{\bibinfo{volume}{73}}
  (\bibinfo{year}{2006}), ISSN \bibinfo{issn}{1550-2368},
  \urlprefix\url{http://dx.doi.org/10.1103/PhysRevD.73.043519}.

\bibitem[{\citenamefont{Floquet}(1883)}]{ASENS_1883_2_12__47_0}
\bibinfo{author}{\bibfnamefont{G.}~\bibnamefont{Floquet}},
  \bibinfo{journal}{Annales scientifiques de l'\'Ecole Normale Sup\'erieure}
  \textbf{\bibinfo{volume}{2e s{\'e}rie, 12}}, \bibinfo{pages}{47}
  (\bibinfo{year}{1883}),
  \urlprefix\url{https://www.numdam.org/articles/10.24033/asens.220/}.

\bibitem[{\citenamefont{{Magnus} and {Winkler}}(1966)}]{MagWink1966}
\bibinfo{author}{\bibfnamefont{W.}~\bibnamefont{{Magnus}}} \bibnamefont{and}
  \bibinfo{author}{\bibfnamefont{S.}~\bibnamefont{{Winkler}}},
  \emph{\bibinfo{title}{{Hill's Equation}}}, vol.~\bibinfo{volume}{xx}
  (\bibinfo{publisher}{Wiley}, \bibinfo{address}{New York},
  \bibinfo{year}{1966}).

\bibitem[{\citenamefont{Adams and Bloch}(2013)}]{Adams_2013}
\bibinfo{author}{\bibfnamefont{F.~C.} \bibnamefont{Adams}} \bibnamefont{and}
  \bibinfo{author}{\bibfnamefont{A.~M.} \bibnamefont{Bloch}},
  \bibinfo{journal}{Journal of Mathematical Physics}
  \textbf{\bibinfo{volume}{54}} (\bibinfo{year}{2013}), ISSN
  \bibinfo{issn}{1089-7658},
  \urlprefix\url{http://dx.doi.org/10.1063/1.4795351}.

\bibitem[{\citenamefont{{Furstenberg} and {Keston}}(1960)}]{Furstenberg1960}
\bibinfo{author}{\bibfnamefont{H.}~\bibnamefont{{Furstenberg}}}
  \bibnamefont{and} \bibinfo{author}{\bibfnamefont{H.}~\bibnamefont{{Keston}}},
  \bibinfo{journal}{Ann. Math. Stat.} \textbf{\bibinfo{volume}{31}},
  \bibinfo{pages}{457} (\bibinfo{year}{1960}).

\bibitem[{\citenamefont{{Furstenberg}}(1963)}]{Furstenberg1963}
\bibinfo{author}{\bibfnamefont{H.}~\bibnamefont{{Furstenberg}}},
  \bibinfo{journal}{Trans. Amer. Math. Soc.} \textbf{\bibinfo{volume}{108}},
  \bibinfo{pages}{377} (\bibinfo{year}{1963}).

\bibitem[{\citenamefont{Adams and Bloch}(2009)}]{AdBloch2009}
\bibinfo{author}{\bibfnamefont{F.~C.} \bibnamefont{Adams}} \bibnamefont{and}
  \bibinfo{author}{\bibfnamefont{A.~M.} \bibnamefont{Bloch}},
  \bibinfo{journal}{J. Math. Phys.} \textbf{\bibinfo{volume}{50}},
  \bibinfo{pages}{073501} (\bibinfo{year}{2009}), \eprint{0906.1954}.

\bibitem[{\citenamefont{{Lima} and {Rahibe}}(1994)}]{LimaRahibe}
\bibinfo{author}{\bibfnamefont{E.}~\bibnamefont{{Lima}}} \bibnamefont{and}
  \bibinfo{author}{\bibfnamefont{M.}~\bibnamefont{{Rahibe}}},
  \bibinfo{journal}{J. Phys. A. Math. Gen.} \textbf{\bibinfo{volume}{27}},
  \bibinfo{pages}{3427} (\bibinfo{year}{1994}).

\bibitem[{\citenamefont{Adams and Bloch}(2010)}]{AdBloch2010}
\bibinfo{author}{\bibfnamefont{F.~C.} \bibnamefont{Adams}} \bibnamefont{and}
  \bibinfo{author}{\bibfnamefont{A.~M.} \bibnamefont{Bloch}},
  \bibinfo{journal}{J. Statist. Phys.} \textbf{\bibinfo{volume}{139}},
  \bibinfo{pages}{139} (\bibinfo{year}{2010}), \eprint{1002.1014}.

\bibitem[{\citenamefont{Dufaux et~al.}(2006)\citenamefont{Dufaux, Felder,
  Kofman, Peloso, and Podolsky}}]{Dufaux_2006}
\bibinfo{author}{\bibfnamefont{J.~F.} \bibnamefont{Dufaux}},
  \bibinfo{author}{\bibfnamefont{G.~N.} \bibnamefont{Felder}},
  \bibinfo{author}{\bibfnamefont{L.}~\bibnamefont{Kofman}},
  \bibinfo{author}{\bibfnamefont{M.}~\bibnamefont{Peloso}}, \bibnamefont{and}
  \bibinfo{author}{\bibfnamefont{D.}~\bibnamefont{Podolsky}},
  \bibinfo{journal}{Journal of Cosmology and Astroparticle Physics}
  \textbf{\bibinfo{volume}{2006}}, \bibinfo{pages}{006–006}
  (\bibinfo{year}{2006}), ISSN \bibinfo{issn}{1475-7516},
  \urlprefix\url{http://dx.doi.org/10.1088/1475-7516/2006/07/006}.

\bibitem[{\citenamefont{Barrowes et~al.}(2024)\citenamefont{Barrowes, Adams,
  Bloch, Giblin, and Watson}}]{paper1}
\bibinfo{author}{\bibfnamefont{L.}~\bibnamefont{Barrowes}},
  \bibinfo{author}{\bibfnamefont{F.~C.} \bibnamefont{Adams}},
  \bibinfo{author}{\bibfnamefont{A.~M.} \bibnamefont{Bloch}},
  \bibinfo{author}{\bibfnamefont{J.~T.} \bibnamefont{Giblin}},
  \bibnamefont{and} \bibinfo{author}{\bibfnamefont{S.}~\bibnamefont{Watson}},
  \bibinfo{journal}{Phys. Rev. D} \textbf{\bibinfo{volume}{110}},
  \bibinfo{pages}{123511} (\bibinfo{year}{2024}),
  \urlprefix\url{https://link.aps.org/doi/10.1103/PhysRevD.110.123511}.

\bibitem[{\citenamefont{Turner}(1983)}]{turner-COs}
\bibinfo{author}{\bibfnamefont{M.~S.} \bibnamefont{Turner}},
  \bibinfo{journal}{Phys. Rev. D} \textbf{\bibinfo{volume}{28}},
  \bibinfo{pages}{1243} (\bibinfo{year}{1983}),
  \urlprefix\url{https://link.aps.org/doi/10.1103/PhysRevD.28.1243}.

\bibitem[{\citenamefont{Braden et~al.}(2010)\citenamefont{Braden, Kofman, and
  Barnaby}}]{Braden:2010wd}
\bibinfo{author}{\bibfnamefont{J.}~\bibnamefont{Braden}},
  \bibinfo{author}{\bibfnamefont{L.}~\bibnamefont{Kofman}}, \bibnamefont{and}
  \bibinfo{author}{\bibfnamefont{N.}~\bibnamefont{Barnaby}},
  \bibinfo{journal}{JCAP} \textbf{\bibinfo{volume}{07}}, \bibinfo{pages}{016}
  (\bibinfo{year}{2010}), \eprint{1005.2196}.

\bibitem[{\citenamefont{Coughlan et~al.}(1983)\citenamefont{Coughlan, Fischler,
  Kolb, Raby, and Ross}}]{COUGHLAN198359}
\bibinfo{author}{\bibfnamefont{G.}~\bibnamefont{Coughlan}},
  \bibinfo{author}{\bibfnamefont{W.}~\bibnamefont{Fischler}},
  \bibinfo{author}{\bibfnamefont{E.~W.} \bibnamefont{Kolb}},
  \bibinfo{author}{\bibfnamefont{S.}~\bibnamefont{Raby}}, \bibnamefont{and}
  \bibinfo{author}{\bibfnamefont{G.}~\bibnamefont{Ross}},
  \bibinfo{journal}{Physics Letters B} \textbf{\bibinfo{volume}{131}},
  \bibinfo{pages}{59} (\bibinfo{year}{1983}), ISSN \bibinfo{issn}{0370-2693},
  \urlprefix\url{https://www.sciencedirect.com/science/article/pii/0370269383910912}.

\bibitem[{\citenamefont{Dimopoulos et~al.}(2008)\citenamefont{Dimopoulos,
  Kachru, McGreevy, and Wacker}}]{Dimopoulos:2005ac}
\bibinfo{author}{\bibfnamefont{S.}~\bibnamefont{Dimopoulos}},
  \bibinfo{author}{\bibfnamefont{S.}~\bibnamefont{Kachru}},
  \bibinfo{author}{\bibfnamefont{J.}~\bibnamefont{McGreevy}}, \bibnamefont{and}
  \bibinfo{author}{\bibfnamefont{J.~G.} \bibnamefont{Wacker}},
  \bibinfo{journal}{JCAP} \textbf{\bibinfo{volume}{0808}}, \bibinfo{pages}{003}
  (\bibinfo{year}{2008}), \eprint{hep-th/0507205}.

\bibitem[{\citenamefont{Gu and Brandenberger}(2018)}]{Gu:2018anb}
\bibinfo{author}{\bibfnamefont{B.-M.} \bibnamefont{Gu}} \bibnamefont{and}
  \bibinfo{author}{\bibfnamefont{R.~H.} \bibnamefont{Brandenberger}},
  \bibinfo{journal}{JCAP} \textbf{\bibinfo{volume}{1805}}, \bibinfo{pages}{017}
  (\bibinfo{year}{2018}), \eprint{1801.08353}.

\bibitem[{\citenamefont{Dasgupta et~al.}(2009)\citenamefont{Dasgupta,
  Brandenberger, Frey, and Lukas}}]{Dasgupta:2004dw}
\bibinfo{author}{\bibfnamefont{K.}~\bibnamefont{Dasgupta}},
  \bibinfo{author}{\bibfnamefont{R.~H.} \bibnamefont{Brandenberger}},
  \bibinfo{author}{\bibfnamefont{A.~R.} \bibnamefont{Frey}}, \bibnamefont{and}
  \bibinfo{author}{\bibfnamefont{A.}~\bibnamefont{Lukas}},
  \bibinfo{journal}{JCAP} \textbf{\bibinfo{volume}{0908}}, \bibinfo{pages}{031}
  (\bibinfo{year}{2009}), \eprint{0812.3620}.

\bibitem[{\citenamefont{Allahverdi et~al.}(2004)\citenamefont{Allahverdi,
  Brandenberger, and Mazumdar}}]{Allahverdi:2004ds}
\bibinfo{author}{\bibfnamefont{R.}~\bibnamefont{Allahverdi}},
  \bibinfo{author}{\bibfnamefont{R.}~\bibnamefont{Brandenberger}},
  \bibnamefont{and} \bibinfo{author}{\bibfnamefont{A.}~\bibnamefont{Mazumdar}},
  \bibinfo{journal}{Phys. Rev. D} \textbf{\bibinfo{volume}{70}},
  \bibinfo{pages}{083535} (\bibinfo{year}{2004}), \eprint{hep-ph/0407230}.

\bibitem[{\citenamefont{Allahverdi et~al.}(2010)\citenamefont{Allahverdi,
  Brandenberger, Cyr-Racine, and Mazumdar}}]{Allahverdi:2010xz}
\bibinfo{author}{\bibfnamefont{R.}~\bibnamefont{Allahverdi}},
  \bibinfo{author}{\bibfnamefont{R.}~\bibnamefont{Brandenberger}},
  \bibinfo{author}{\bibfnamefont{F.-Y.} \bibnamefont{Cyr-Racine}},
  \bibnamefont{and} \bibinfo{author}{\bibfnamefont{A.}~\bibnamefont{Mazumdar}},
  \bibinfo{journal}{Ann. Rev. Nucl. Part. Sci.} \textbf{\bibinfo{volume}{60}},
  \bibinfo{pages}{27} (\bibinfo{year}{2010}), \eprint{1001.2600}.

\end{thebibliography}

\end{document}